# GENERATIVE LSTM MODELS AND ASSET HIERARCHY CREATION IN INDUSTRIAL FACILITIES

**Morgen Pronk**
MIT, SDM Master's Student
Cambridge, MA

## ABSTRACT

*In the evolving field of maintenance and reliability engineering, the organization of equipment into hierarchical structures presents both a challenge and a necessity, directly impacting the operational integrity of industrial facilities. This paper introduces an innovative approach employing machine learning, specifically Long Short-Term Memory (LSTM) models, to automate and enhance the creation and management of these hierarchies. By adapting techniques commonly used in natural language processing, the study explores the potential of LSTM models to interpret and predict relationships within equipment tags, offering a novel perspective on understanding facility design.*

*Our methodology involved character-wise tokenization of asset tags from approximately 29,000 entries across 50 upstream oil and gas facilities, followed by modeling these sequences using an LSTM-based recurrent neural network. The model's architecture capitalizes on LSTM's ability to learn long-term dependencies, facilitating the prediction of hierarchical relationships and contextual understanding of equipment tags.*

*The results reveal that the LSTM model effectively predicts character-wise probabilities and discerns hierarchical relationships among tags. While it demonstrates a high accuracy level in understanding the context and relationship of equipment tags, particularly in cases with complex inter-equipment dependencies, limitations were observed in consistently predicting certain numeric elements within tags. Despite these constraints, the model substantially reduces the workload for reliability engineers in organizing equipment hierarchies and shows promise for broader applications in engineering design.*

*This study sets a precedent for the use of LSTM models in industrial applications beyond traditional language processing, highlighting future opportunities for integrating more advanced techniques like transformer models, especially with larger datasets. The findings of this research pave the way for further exploration into the potential of machine learning in enhancing operational efficiency and design capabilities within the realm of reliability engineering.*

## 1. INTRODUCTION

In the realm of maintenance and reliability engineering, efficiently managing equipment hierarchies stands as a significant challenge, with direct implications for the operational integrity of facilities. This study introduces a novel approach utilizing machine learning techniques to not only assist in automating the creation of these hierarchies, thereby reducing the workload on engineers, but also to extend its utility in assisting engineering design. Inspired by the capabilities of Large Language Models, which generate logical statements based on sequential word probability calculations, the hypothesis of this paper posits that a similar machine learning model could discern inter-equipment relationships and infer facility design principles by learning from existing equipment hierarchies. Remarkably, the model demonstrates the potential to generate tags for previously unseen equipment, suggesting an adaptive, predictive capacity that could redefine equipment management and design methodologies in engineering.

## 2. BACKGROUND

In the realm of industrial processing, notably in sectors like upstream oil and gas and chemical manufacturing, the organization of assets, including equipment, into hierarchies is a critical practice for maintenance and reliability. These hierarchies, often structured in accordance with ISO 14224 standards, play a pivotal role in reliability calculations and in illustrating the dependencies and relationships among various assets. While the ISO 14224 standards provide a foundational framework, it is commonplace for companies to tailor these standards to their specific operational needs.

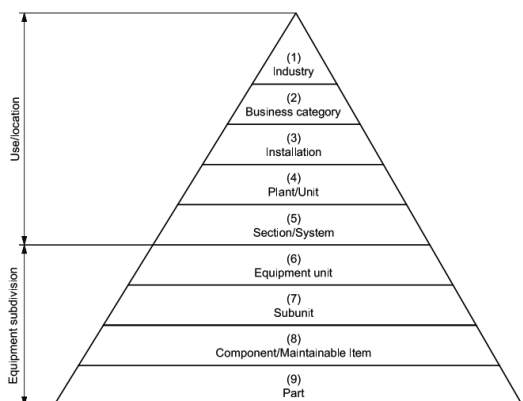

*Figure 1: ISO 14224 general hierarchy*

## 2.1. UNDERSTANDING THE TAGGING SYSTEM

At the core of these asset hierarchies lies a sophisticated tagging system. Unlike serial numbers that identify the physical entity of an asset, these tags, often referred to as "functional locations," are primarily concerned with the asset's function and its relative position within the facility. These tags are integral to a Computerized Maintenance Management System (CMMS), where they serve as vital identifiers encoding information about the facility's design and expected asset arrangement.

Each tag within this system comprises various alphanumeric codes, encapsulating key information about the facility and its design. Below are examples of the components that commonly make up upstream oil and gas facilities and their tags:

- Project Code: Identifies the major project to which the facility belongs, with each project code implying a different type of facility.
- Area Code: Alphanumeric representation of the facility's location, also linked to the type of oil, gas, and by-products produced in that area.
- Train Number: A numeric indicator of the process train within the facility to which the equipment belongs.
- Facility Type: Alphanumeric code denoting the nature of the facility, such as 'W' followed by three integers for well sites.
- Unit Identification: Numeric code correlating to the specific system within the facility, like '17' for compression systems.

An example of an equipment level functional location tag in an oil and gas facility is shown in Figure 2.

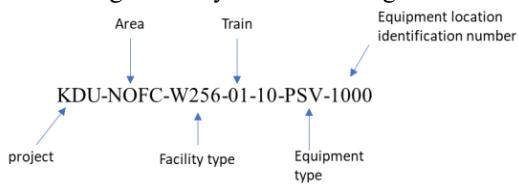

*Figure 2: Example of an equipment tag and contextual information it contains*

In the context illustrated by Figure 2, a reliability or maintenance engineer can decipher a wealth of information from the tag's structure and its embedded codes. In example shown in Figure 2, the tag indicates a specific piece of equipment: the primary, or first, pressure safety valve (PSV) in the facility, denoted by the code '1000' (with '1001' representing the second PSV). This PSV is part of the inlet gas system (unit identification 10), located on train 01, at the well site W256 in the North of Fox Creek area. It is associated with the Kaybob Duvernay non-exploration project, as indicated by the code 'KDU' (as opposed to 'KDX').

This tag's structure and the relationships among its components reveal critical information about the facility's design and operational setup. In this specific instance, the well site W256, under the KDU project in the NOFC area, employs a PSV as the safety mechanism for the wellhead's pressure overprotection. Hence, the parent of this tag would be identified as KDU-NOFC-W256-01-10-W-1000. To accurately predict such hierarchical relationships, a model would require an in-depth understanding of the facility's design. It needs to recognize, for instance, that in this particular design, PSVs are used to protect wellheads, and that this specific PSV is dedicated to the wellhead, rather than to another component like the separator within the same facility. This nuanced understanding of the facility's design and the functional implication of its equipment is what the project aims to leverage through machine learning.

There is significance immediate practical value in this project as well, especially in the context of establishing and maintaining industrial facilities. Whether it involves the completion of a new facility or implementing modifications in an existing one, the upkeep of these tags is crucial. For maintenance and reliability engineers, a primary and time-intensive responsibility is the systematic organization of all equipment tags into their appropriate hierarchical structures. This involves meticulously identifying the 'superior' functional locations for each tag, effectively mapping out the parent-child relationships within the facility's equipment network. In the case of a new facility, this task is particularly daunting, often requiring several weeks, if not months, to complete.

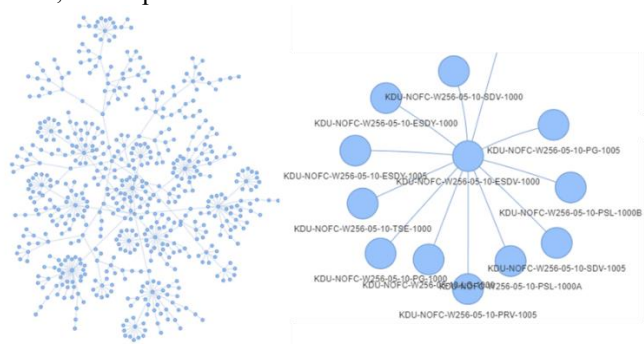

*Figure 3: Graph representation of equipment in upstream gas facility. Right shows the parent-child relationships between the equipment. Left shows a zoomed in view of one of the clusters with the Functional location tags*

This process is not only laborious but also critical for the efficient and safe operation of the facility. Properly organized hierarchies ensure accurate tracking and maintenance of equipment, which is essential for operational reliability and compliance with safety standards. The potential for automating this process through machine learning offers a promising avenue to significantly reduce the time and effort involved, while possibly enhancing the accuracy and consistency of the hierarchical structuring.

## 3. METHOD

The genesis of this project lies in the innovative domain of Transformer-based modeling. This approach, a cornerstone in the field of natural language processing, involves generating responses to queries by calculating and selecting the most probable subsequent word in a sequence. This selection is based on extensive training on large text datasets. Although it might initially seem counterintuitive, understanding the probability of the next word has proven to be a remarkably effective method for comprehending human language and its underlying knowledge structures. Transformer architectures excel in capturing contextual nuances and intricate relationships within language. Drawing inspiration from this, the aim of this paper is to develop a model that, by predicting the characters in an equipment tag sequentially, can similarly derive in-depth insights into the design of industrial facilities. This is somewhat analogous to Large-language models understanding much about human knowledge though sequentially predicting the most probable words.

Despite drawing inspiration from Transformer models, the methodology for this project pivots towards utilizing a Long Short-Term Memory (LSTM) Recurrent Neural Network. Although architecturally distinct from Transformers, LSTMs share commonalities such as being neural network-based, relying on backpropagation and gradient-based optimization, and their aptitude for handling sequential data. Both are adept in applications like text generation and machine translation, where understanding the sequence and context of input data is crucial. However, LSTMs and Transformers learn data representations differently: LSTMs leverage their recurrent nature and memory cells, while Transformers utilize self-attention mechanisms to assess the relative importance of various parts of the input sequence.

The decision to employ LSTM in this project was influenced by several critical factors. Firstly, the dataset used is relatively small, comprising approximately 29,000 tags from around 50 different facilities. LSTMs have been observed to outperform Transformers in scenarios with limited data (Ezen-Can, 2020). The smaller dataset implies fewer instances of tags in specific scenarios, which suits the capabilities of LSTMs. Secondly, computational resources were a significant consideration. LSTMs are generally more computationally efficient than Transformers when dealing with small to moderate-sized models and datasets (Hahn, 2020). Given the resource constraints of this project LSTMs were chosen as a more suitable choice for the objectives.

### 3.1. LSTM MODEL

This section provides an overview of Long Short-Term Memory (LSTM) Models and Recurrent Neural Networks (RNN).

Traditional feedforward neural networks, while effective in various applications, have a notable limitation: they cannot use information from previous outputs to inform future predictions without undergoing retraining. This limitation is especially critical in tasks where sequential data and historical context are important. To overcome this challenge, Recurrent Neural Networks (RNNs) were developed. RNNs introduce loops in their architecture, allowing information to pass from one step of the network to the next. This approach effectively enables the network to maintain a memory of prior inputs, facilitating sequential predictions. Figure 4 illustrates how RNNs loop outputs from previous predictions back into the network.

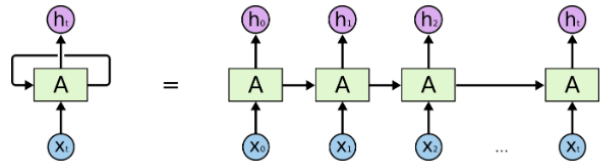

*Figure 4: RNN pass outputs from previous predictions back to the neural network in a loop. (Olah, 2015)*

RNNs still have a problem; they face difficulties in bridging long temporal gaps between relevant pieces of information. This is where Long Short-Term Memory (LSTM) models, a special kind of RNN, become advantageous. LSTMs are designed to capture long-term dependencies and can be visualized as a chain of repeating modules, each passing information to the next.

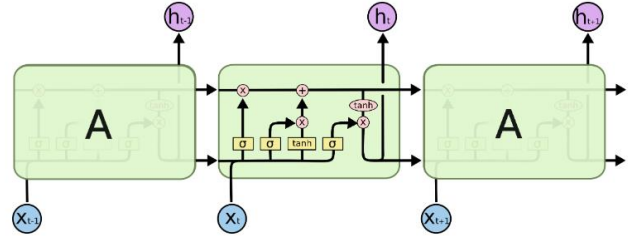

*Figure 5: Chain Structure of a LSTM model (Olah, 2015)*

Each of these modules has four interacting neural network layers and a cell state that carries information through to each module. This cell state is the main concept that allows LSTM models to develop long-term dependencies.

The first of the neural network layers passes the input, $x_t$, through a sigmoid function that determines what information it will cause the cell state, $C_{t-1}$, to forget using point wise multiplication. This is called the "forget gate layer".

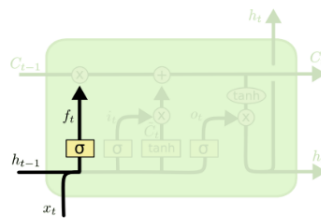

$$f_t = \sigma\left(W_f \cdot [h_{t-1}, x_t] + b_f\right)$$

*Figure 6: Forget gate layer of a LSTM model. (Olah, 2015)*

The next step determines what new information to store in the cell state by using another sigmoid function, $i_t$. This is called the "input gate layer". Unlike the previous step this does not go directly to the cell state and instead combines using pointwise multiplication with the output of a tanh layer, $\tilde{C}_t$. This determines what new information to add to the cell state to pass to the next module and prediction.

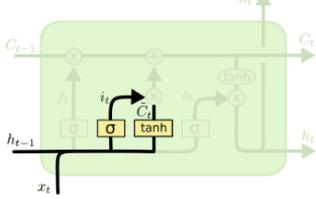

$$i_t = \sigma(W_i \cdot [h_{t-1}, x_t] + b_i)$$
$$\tilde{C}_t = \tanh(W_C \cdot [h_{t-1}, x_t] + b_C)$$

*Figure 7: Input Gate Layer of a LSTM model. (Olah, 2015)*

The result is a cell state that is passed along to the next module and allows relevant information to be passed throughout the entire chain. This cell state is represented by the equation below:

$$C_t = f_t C_{t-1} + i_t \tilde{C}_t$$

*Equation 1: Final Cell State*

The final layer is an output prediction layer that combines a sigmoid function and a tanh-filtered version of the cell state. This layer decides what part of the current sequence should contribute to the current prediction and is illustrated by Figure 8.

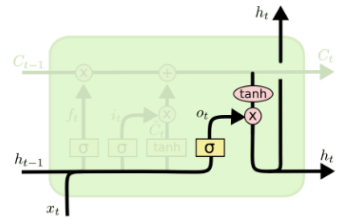

$$o_t = \sigma(W_o [h_{t-1}, x_t] + b_o)$$
$$h_t = o_t * \tanh(C_t)$$

*Figure 8: Output layer for a single module of a LSTM module. (Olah, 2015)*

Through this process the model can retain and change information about long term dependencies of sequential information and use that to predict elements of sequences.

### 3.2. MODEL ARCHITECTURE AND DESIGN

In this project, tags were modeled as a sequence of characters and the model was trained to understand the dependencies between the characters but also the hierarchical relationships between tags. The architecture used to do this is primarily a LSTM-based RNN with a fully connected SoftMax output layer illustrated by Figure 9.

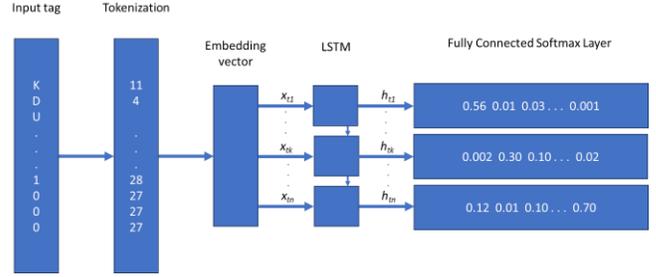

*Figure 9: Architecture used to predict parent-child relations between equipment using character-wise prediction.*

The initial stage of the modeling process involves transforming input tags into tokenized vectors. We chose a vector length of 40, an arbitrary but sufficiently large number to encompass the longest tag in our dataset. This uniform length is crucial to ensure consistent input size for the model. Tags shorter than 40 characters are padded with zeros, which represent empty strings in the tokenized format.

Our approach focuses on character-wise sequential prediction, thereby opting for character-wise tokenization. This method is particularly beneficial as it allows the model to generalize across various tagging conventions and predict tags for equipment not included in the training dataset. It also facilitates the generation of new tags. Had we tokenized based on known elements within the tags, such as specific codes like "KDU", "PSV", or "W256", the model would be limited to the scope of its training data. In such a scenario, encountering new, unseen codes would necessitate retraining the entire model, as it would not inherently know how to tokenize these novel inputs.

Following tokenization, the model employs an embedding layer to transform the tokenized vectors' sparse integers into dense vectors. This transformation is key for effectively capturing the nuances in the data. The dense vectors are then fed into LSTM layers, which are adept at learning the contexts and dependencies between individual characters within the tags. These LSTM layers generate latent space representations of each tag, encapsulating the learned patterns and relationships.

The next phase involves a fully connected layer equipped with a SoftMax function. This layer's role is to convert the latent space representations into a probability matrix. Each element in this matrix indicates the likelihood of a character being the correct next character in the tag's sequence.

To fine-tune the model's ability to discern these probabilities, it was trained on a dataset comprising approximately 29,000 tags from 50 different upstream oil and gas facilities. The training process aimed to establish a robust mapping between input tags and their corresponding output probabilities of characters and their sequences for the parent tags. A modified cross-entropy loss function was utilized during this training phase to optimize the model's accuracy. The formula for this loss function is presented below:

$$H(y, p) = \sum_i -y_i \log(p_i) + a(1 - y_i) p_i$$

Where:

- $y_i$ is the true label for class $i$ (1 for the true class, 0 for all others).
- $p_i$ is the predicted probability for class $i$.
- $a$ is a scaling parameter to increase the penalty for high probabilities.

*Equation 2: Loss function used for model training*

In this project, each character within a tag is represented as a unique class token, and cross-entropy serves as the loss function. Cross-entropy is a standard choice for classification problems, as it effectively quantifies the difference between the predicted and actual probability distributions. In this case this would be the probabilities of each character in the tag, based on the characters before it. The tag is reversed before being given to the LSTM, as the most specific and unique codes – and therefore the important characters - tend to come at the end of the tag. As LSTMs are sequential and not bi-directional, like transformer models such as BERT, this resulted in a significant increase in performance. During the training phase, we discovered that imposing an additional penalty for incorrectly high probabilities — those not aligning with the actual character in a specific sequence segment — notably enhanced the model's performance. This adjustment also led to a significant reduction in the occurrence of incorrect characters within the padded sections of the tags.

For making predictions with the trained model, it first estimates the probability of each character in every position of a given input tag. These probabilities provide a level of interpretability for the model's outputs. The character corresponding to the highest probability token is selected and then transformed back into its original form using an inverse tokenizer function.

To evaluate and monitor the model's effectiveness the Levenshtein distance, also known as edit distance, was used. This metric was used both during the training process and for post-training performance assessment. Detailed insights into this aspect of the model's evaluation are discussed in Section 4.

Finally, fine-tuning a pre-trained LLM was considered as an alternative to using an LSTM model, but was ultimately not used for several reasons. First, this task involves predicting specific structured character sequences that do not follow similar structure or conventions of English. Most LLMs that are available for use excel in understanding and generating human-like text, not predicting structured, non-linguistic character sequences. To fine tune a model on such a fundamentally different structure would likely require much more training data than available or generate undesirable results. Second, finetuning an LLM requires much more computation power than training a LSTM based model like ours. With the limited resources available for the project, and to the typical reliability/maintenance engineer, the simplicity of the LSTM approach was considered a significant advantage.

## 4. RESULTS AND DISCUSSION

The model's results demonstrate its capability to effectively predict character-wise probabilities and discern the hierarchical relationships between tags, thereby offering insights into the design of the facilities. Notably, the model exhibits strong performance in predicting tags associated with equipment, components, and parts. To evaluate the model's practical applicability and effectiveness, a complete site was reserved exclusively for testing purposes. This approach, diverging from the conventional method of randomly selecting tags for testing, allowed for a comprehensive assessment of the model's performance across an entire site. This testing methodology aligns closely with the real-world scenario of assisting a reliability engineer in establishing the hierarchy and relationships among various pieces of equipment. To showcase the model's understanding of these relationships, a selection of similar tags and equipment from the test site has been compiled and presented in a tabular format below:

| Child Tag | Parent Tag | Prediction | Differences | Levenshtein Distance |
|---|---|---|---|---|
| KDU-NOFC-W256-01-10-PSV-1000 | KDU-NOFC-W256-01-10-W-1000 | KDU-NOFC-W05--01-10-W-1000 | KDU-NOFC-W*5*-01-10-W-1000 | 2 |
| KDU-NOFC-W256-01-10-PSV-1055 | KDU-NOFC-W256-01-10-V-1055 | KDU-NOFC-W059-01-10-V-1055 | KDU-NOFC-W*5*-01-10-V-1055 | 2 |
| KDU-NOFC-W256-01-55-PSV-5570 | KDU-NOFC-W256-01-55-P-5570 | KDU-NOFC-W052-01-55-P-5570 | KDU-NOFC-W*5*-01-55-P-5570 | 2 |
| KDU-NOFC-W256-01-10-PSV-1060 | KDU-NOFC-W256-01-10-V-1060 | KDU-NOFC-W05--0--10-V-1060 | KDU-NOFC-W*5*-0*-10-V-1060 | 3 |
| KDU-NOFC-W256-01-10-TIT-1063 | KDU-NOFC-W256-01-10-V-1060 | KDU-NOFC-W055-01-10-V-1060 | KDU-NOFC-W*5*-01-10-V-1060 | 2 |
| KDU-NOFC-W256-01-10-TIT-1011 | KDU-NOFC-W256-01-10-W-1000 | KDU-NOFC-W055-01-10-W-1010 | KDU-NOFC-W*5*-01-10-W-10*0 | 3 |
| KDU-NOFC-W256-01-10-TIT-1000 | KDU-NOFC-W256-01-10-W-1000 | KDU-NOFC-W015-01-10-W-1000 | KDU-NOFC-W***-01-10-W-1000 | 3 |
| KDU-NOFC-W256-02-10-TIT-1011 | KDU-NOFC-W256-02-10-W-1000 | KDU-NOFC-W052-01-10-W-1010 | KDU-NOFC-W*5*-0*-10-W-10*0 | 4 |
| KDU-NOFC-W256-01-10-TIT-1057 | KDU-NOFC-W256-01-10-V-1055 | KDU-NOFC-W015-01-10-V-1057 | KDU-NOFC-W***-01-10-V-105* | 4 |
| KDU-NOFC-W256-01-10-AE-1066 | KDU-NOFC-W256-01-10-AIT-1065 | KDU-NOFC-W145-01-10-AIT-1065 | KDU-NOFC-W***-01-10-AIT-1065 | 3 |

*Figure 10: Results of predicted relationships between equipment using a LSTM and character-wise tokenization.*

Figure 10 shows the model input (Child Tag), the ground truth (Parent Tag), the model prediction (Prediction) and highlights where the model did not quite get the tag right (Differences) as well as the Levenshtein distance. This metric evaluates the similarity between character strings by counting the number of single-character edits needed to transform one string into another. The first important observation is that despite the equipment all being of similar types, Pressure Safety Valves (PSV) and Temperature Indicator/Transmitter (TIT), the model can learn that the relationship between the equipment and its context. For example, the first tag "KDU-NOFC-W256-01-10-PSV-1000" protects the well head and therefore is predicted to be related the tag for the wellhead, "KDU-NOFC-W256-01-10-W-1000". However, a different PSV, "KDU-NOFC-W256-01-55-PSV-5570", is predicted to be related to the methanol pump "KDU-NOFC-W256-01-55-P-5570", which is correct as it the safety pressure valve for that pump. The model is even able to get these tags correct even when there are less direct similarities between the tags. The equipment tag "KDU-NOFC-W256-01-10-AE-1066" is an example of this, where it was predicted to be related to the tag "KDU-NOFC-W256-01-10-AIT-1065". This shows that the model can understand the context of the tag and make correct predictions of similar equipment by separate contexts and not just overfitting to certain patterns.

However, the model is not, admittedly, perfect. In Figure 10, a pattern of mislabels are apparent. The model consistently mislabels characters associated with the well number. One potential explanation for this may be that for this project and area (KDU-NOFC) all the wells share a common design, due to being built around the same time and for similar reservoirs in a common area, and so remembering the well pad number is not relevant for the model and the LSTM likely chooses to forget this information in the forget gate layer, and is not passed on into the cell state. In other words, it could be that this part of the tag is not significant in determining the relations of the equipment. Regardless of if the tag was W256 or W059 both would have negligibly different probabilities according to the model because the relationship between equipment it is predicting is the same across all wells in this area.

Another interesting finding is how the model performs not only at the equipment levels of the hierarchy, but in other levels and across levels. Figure 11 shows how the model performs across the hierarchy by following a chain of parent-child tag relationships:

| | Child Tag | Parent Tag | Prediction | Differences | Levenshtein Distance |
|---|---|---|---|---|---|
| 1 | KDU-NOFC-W256-02-10-ESDY-1000 | KDU-NOFC-W256-02-10-ESDV-1000 | KDU-NOFC-W159-02-10-ESDV-1000 | KDU-NOFC-W*5*-02-10-ESDV-1000 | 2 |
| 2 | KDU-NOFC-W256-02-10-ESDV-1000 | KDU-NOFC-W256-02-10-W-1000 | KDU-NOFC-W142-01-10-W-1000 | KDU-NOFC-W***-0*-10-W-1000 | 4 |
| 3 | KDU-NOFC-W256-02-10-W-1000 | KDU-W256-0010-0122-WELL | KDU-W199-0010-0122-1000 | KDU-W***-0010-0122-**** | 7 |
| 4 | KDU-W256-0010-0122-WELL | KDU-W256-0010-0122 | KDU-W115-0000-1100 | KDU-W***-00*0-*1** | 7 |
| 5 | KDU-W256-0010-0122 | KDU-W256-0010 | KDU-W115-0120 | KDU-W***-0**0 | 5 |
| 6 | KDU-W256-0010 | KDU-W256 | KDU-W010 | KDU-W*** | 3 |
| 7 | KDU-W256 | KDU | 255 | *** | 3 |

*Figure 11: Tabular summary of predictions of the model down a chain of the hierarchy.*

As can be seen by Figure 11, the model is able to make relational predictions well between the equipment levels of the hierarchy (rows 1 and 2), gets the structure of the tag correct but many of the numeric contents wrong for transitions between the equipment and the section tags (row 3). Withing the section tags (row 4 – 7) it gets the structure of the tag correct, but the contents of the tag incorrect. Although not ideal, for the practical application where a reliability engineer would be creating a hierarchy, these sections follow more predictable patterns and are easier to correct, and so this model would still be extremely useful. Fortunately, the model performs the best in the section that is the most valuable, the equipment levels, where equipment is more diverse, more tags exist, and relationships are less intuitive and require engineering design contextual knowledge.

In general, the model performs very well, and some summary metrics are shown in Figure 12.

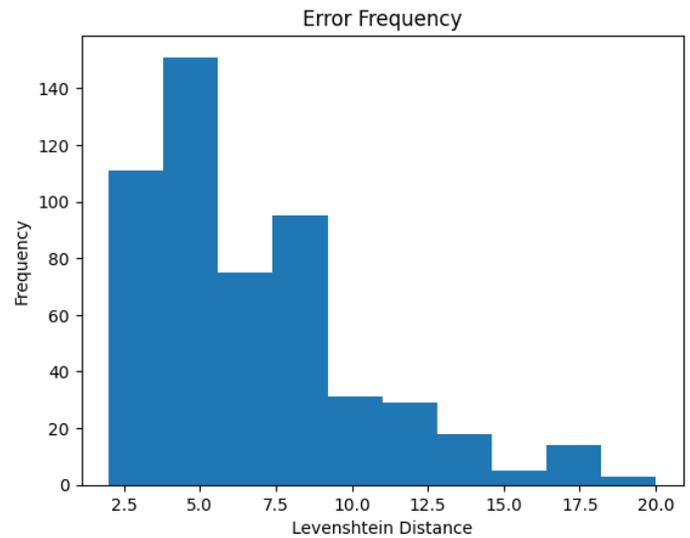

*Figure 12: Error frequency for a full facility test. Levenshtein distance represents the number of character "edits" that would need to be made to make the predicted tag the same as the ground-truth.*

The average Levenshtein distance of the tags for the test dataset was approximately 5.3. This means that on average, there are 5 characters that need to be edited to change the tag into the ground truth value. This is very good considering that a significant number of mistakes occur in characters, such as the well ID or train number, and often the relation between equipment is properly predicted. This level of performance is sufficient for significantly decreasing the amount of work a maintenance or reliability engineer would have to do, and although not capable of being full automated without mistakes, the model is able to understand much of the facility design through understanding the tag structure and the hierarchy relationships.

## 5.  LIMITATIONS AND FUTURE WORK

The model performs well in general, however this must be taken in context of the small data set that it was trained on. Typically, Natural Language Processing models, like ChatGPT, train on datasets consisting millions to billions of tokens (Hoffmann, 2022). Our dataset consists of about 29000 tags, which only represents about 50 relatively small upstream oil and gas processing plants. This is particularly important because it means that the model only has about 50 different facility contexts to differentiate between and to learn relationships from. Having more, and varied data, may help the model be able to make better predictions, or may open possibilities for using methods like transformer models, which are commonly done for machine translation applications.

Other ideas for improving the performance of the model are, training a classifier that could first properly classify the hierarchy of the tag, and then apply specific LSTM models that were trained only on relationships within that level of the hierarchy as it seems the current model performs better if it is dealing with tags of all the same type, like equipment-to-equipment relationships. Another possible improvement is to constrain the model with rules about the tags themselves. This was avoided initially as it assumes rules about the tags must be known a priori. The current model does not make any of these assumptions and therefore should be more generalizable regardless of the specific organizations tagging conventions. Whether this is true or not would have to be tested by either synthesizing different tagging standards for the facility data in the dataset, or looking at other organizations or projects that may use significantly different conventions and is another area that would require some future work.

## 6. CONCLUSION

This project has successfully demonstrated the potential of an LSTM-based model in leveraging unconventional data sources, such as equipment tags, to gain insights into facility design. This approach is particularly valuable in its practical application, offering substantial improvements to the tasks typically performed by reliability and maintenance engineers in processing facilities. While the model does not yet enable full automation in constructing equipment hierarchies for Computerized Maintenance Management Systems (CMMS) and reliability systems, its proficiency in assisting and augmenting this process is nonetheless noteworthy. The model exhibits a commendable accuracy level, with relatively few character-wise errors that are easily identifiable by human reviewers. More importantly, it accurately captures essential equipment relationships, which is crucial for effective facility management.

Despite these achievements, there are several avenues for further enhancement of the model. Future work could explore the integration of other techniques, such as transformer models, particularly in the context of larger datasets. The potential improvements and expanded applications that these advancements could bring to the field of facility management and reliability engineering present exciting opportunities for continued research and development.

## 7. REFERENCES

Ezen-Can, A. (2020). A comparison of LSTM and BERT for Small Corpus. *arXiv*.

Hahn, M. (2020). Theoretical Limitations of Self-attention in Neural Sequence Models. *Association of Computations Linguistics*, 8:156-171.

Hoffmann, J. (2022). Training Compute-Optimal Large Language Models. *arXiv*, 3.

Olah, C. (2015, August 27). *Understanding LSTM Networks.* Retrieved from https://colah.github.io/posts/2015-08-Understanding-LSTMs/